\documentclass[linenumbers]{aastex631}

\usepackage{amssymb}
\usepackage{amsmath}
\usepackage{epsfig}
\usepackage{epstopdf}
\usepackage{comment}
\usepackage{graphicx,wrapfig}
\usepackage{color}
\usepackage{soul}
\usepackage{tabularx,colortbl}
\usepackage[version=3]{mhchem}
\usepackage{placeins}

\definecolor{lavender}{rgb}{0.75, 0.58, 0.89}

\newcolumntype{M}[1]{>{\centering\arraybackslash}p{#1}}
\newcolumntype{P}[1]{>{\raggedright\arraybackslash}p{#1}}

\newcommand{\prt}{\partial}
\newcommand{\bs}{\mbox{\boldmath $s$}}
\newcommand{\baa}{\mbox{\boldmath $a$}}

\newcommand{\bu}{\mbox{\boldmath $u$}}
\newcommand{\bw}{\mbox{\boldmath $w$}}

\newcommand{\bvv}{\mbox{\boldmath $v$}}

\newcommand{\bxi}{\mathbf{y}}

\newcommand{\balpha}{\mbox{\boldmath $\alpha$}}
\newcommand{\lp}{\left(}
\newcommand{\rp}{\right)}

\newcommand{\non}{\nonumber}

\newif\ifcomments % with (or without) personalized/colored comments
\commentsfalse
\commentstrue % put % in front if you want comments not to be shown  

\definecolor{Green}{rgb}{0,0.5,0}

\modulolinenumbers[5]
\definecolor{lightgray}{gray}{0.9}
\definecolor{amethyst}{rgb}{0.8, 0.0, 0.8}
\definecolor{aogreen}{rgb}{0.01, 0.75, 0.24}

\shorttitle{Skeletal Kinetics Reduction of Reaction Networks}
\shortauthors{Nouri et al.}

\begin{document}

\nolinenumbers

\title{Skeletal Kinetics Reduction for Astrophysical  Reaction Networks}

\author[0000-0001-7390-5212]{A.G. Nouri}
\affiliation{Department of Mechanical Engineering and Materials Science, University of Pittsburgh, Pittsburgh, PA 15261, USA}

\author[0000-0001-5261-8994]{Y. Liu}
\affiliation{Department of Mechanical Engineering and Materials Science, University of Pittsburgh, Pittsburgh, PA 15261, USA}

\author[0000-0002-9557-5768]{P. Givi}
\affiliation{Department of Mechanical Engineering and Materials Science, University of Pittsburgh, Pittsburgh, PA 15261, USA}

\author[0000-0002-6318-2265]{H. Babaee}
\affiliation{Department of Mechanical Engineering and Materials Science, University of Pittsburgh, Pittsburgh, PA 15261, USA}

\author[0000-0003-2367-1547]{D. Livescu}
\affiliation{Los Alamos National Laboratory, Los Alamos, NM 87544, USA}

%\author[0000-0002-0474-159X]{F.X. Timmes}
%\affiliation{School of Earth and Space Exploration, Arizona State University, Tempe, AZ 85287, USA}
%\affiliation{Joint Institute for Nuclear Astrophysics - Center for  the  Evolution of the Elements, USA}

\begin{abstract}
\nolinenumbers
A novel methodology is developed to extract accurate skeletal reaction models for nuclear combustion. Local sensitivities of isotope mass fractions with respect to reaction rates are modeled based on the forced optimally time-dependent (f-OTD) scheme. These sensitivities are then analyzed temporally to generate skeletal models. The methodology is demonstrated by conducting skeletal reduction of %hydrostatic
constant density and temperature burning of carbon and oxygen relevant to SNe Ia. The 495-isotopes Torch model is chosen as the detailed reaction network. A  map of maximum production of $^{56}\text{Ni}$ in SNe Ia is produced for different temperatures, densities, and proton to neutron ratios. The f-OTD simulations and the sensitivity analyses are then performed with initial conditions from this map.  A series of skeletal models  are derived  and their performances are assessed by comparison against currently existing skeletal models. Previous models have been constructed intuitively by assuming the dominance of $\alpha$-chain reactions. The comparison of the newly generated skeletal models against previous models is based on the predicted energy release and  $^{44}\text{Ti}$ and $^{56}\text{Ni}$ abundances by each model. The consequences of $\mathtt{y}_e \neq 0.5$ in the initial composition are also explored where $\mathtt{y}_e$ is the electron fraction. The simulated results show that $^{56}\text{Ni}$ production decreases by decreasing  $\mathtt{y}_e$ as expected, and that the  $^{43}\text{Sc}$ is a key isotope in  proton and neutron channels toward $^{56}\text{Ni}$ production. It is shown that an f-OTD skeletal model with 150 isotopes can accurately predict the $^{56}\text{Ni}$ abundance in SNe Ia for $\mathtt{y}_e \lesssim 0.5$ initial conditions.

\end{abstract}


\section{Introduction} \label{sec:intro}
Direct implementation of detailed reaction networks (RNs) containing many isotopes and reactions in hydrodynamic flow solvers is computationally very expensive. In most cases, it is unavoidable to use efficient reaction kinetics models to conduct large-scale hydrodynamic simulations pertaining to astrophysical explosions, such as Type Ia supernovae (SNe Ia). These reaction kinetics models, which are usually extracted from detailed reaction networks, should reasonably estimate the released energy and isotope abundances. Integration of the ordinary differential equations representing the abundance levels of a set of isotopes of reacting nuclei in the continuum limit serves two functions in stellar models. The primary function of hydrodynamics is to provide the magnitude and sign of the nuclear energy generation rate \citep{Weaver1978,BBM19,AG20}. This is usually the largest energy source in regions conducive to nuclear reactions, and its accurate determination is essential for stellar models. These models usually require accurate predictions of the energy generated by nuclear burning over a wide range of temperatures, densities, and compositions~\citep{Timmes99,THW00,RS18}. The other function is to  describe the evolution of the composition.  In some stellar events, the isotopic abundances themselves are of primary interest for understanding the origin and evolution of the chemical elements \citep{Pagel09,Matteucci12,KKL20}. Moreover, matching observational evidence of certain isotopes, \textit{e.g.} $^{56}\text{Ni}$ in supernova light curves, gives confidence in the underlying computational model \citep{ST17,BVK22}.

Thousands of isotopes can participate in a reaction network during  a stellar phenomenon \citep{Wanajo13,NTT15,FFKLDR17,LR17,Psaltis22}. Accurate predictions of the nuclear energy generation rate and the composition changes in such RNs is computationally expensive. The largest block of memory in a stellar model  is usually used for storing the isotopic abundances of every computational cell at all time steps. Even with modern methods for solving  stiff systems of ordinary differential equations, integration of the evolution equations of the isotopic abundances dominates the total cost of a stellar model when the number of  isotopes evolved is $\mathcal{O}$(100) \citep{Arnett96,NGL19}. To decrease the computational cost, one has to make a choice between having fewer isotopes (order reduction) or less spatial resolution (or mass resolution). The general response to this trade-off has been the order reduction by using simplified RNs within hydrodynamic solvers to calculate an approximate energy generation rate and isotope mass fractions during stellar explosions \citep{RS18}. As a post-processing step, the detailed nuclear composition of the ejecta is computed using a large RN by employing Lagrangian tracer particles \citep{TNY86,TMTCB16,LN18,BBM19,SP23}. These particles represent passive mass elements and can be evolved in situ, within the simulation, with time steps dictated by the hydrodynamic solvers, or off-site, using an additional recontruction step based on the snapshot data \citep{Setal23}. Current large simulations can use at least $10^6$ tracer particles to calculate detailed 3D spatial composition \citep{SP23}.

Order reduction techniques for nuclear reaction networks require extensive experience and expertise. For example, common $\alpha$-chain RNs (with 13/19/21 isotopes \citep{MESA,Paxton15}) contain a minimal set of nuclei to approximate the energy generation rate for stellar simulations of SNe Ia \citep{MESA}. These RNs are optimized by decades of shared knowledge \citep{YTKTNU21}. Order reduction techniques for nuclear RNs also include  explicit asymptotic  \citep{Guidry13a}, quasi-steady-state~(\citep{MOV00,Guidry13b}), and quasi-equilibrium~(\citep{BCF68,WAC73,Khokhlov81,MKC98,HKWT98,THW00,HPFT07,KK20}) methods. However, these RNs, while fast and lightweight, are rigid with respect to adding or removing isotopes \citep{MESA}.

In chemical combustion, recent advances in data-driven techniques and sensitivity analysis have opened the possibility of significantly enhancing the efficiency and flexibility of generating reduced RNs \citep{LL09}. Recent skeletal models, \textit{i.e.} optimized subsets of detailed reaction networks, can be prepared in an optimized and automated manner, with consistent accuracy throughout the evolution of the network, and can be adapted based on the availability of computing resources. Utilizing such capabilities for nuclear combustion can address certain issues regarding the model reduction of nuclear RNs, \textit{e.g.} scenarios where the experience required to generate quality reduced models might be lacking. There is a history of utilizing sensitivity analysis in nuclear combustion. This includes studies of the Big Bang~\citep{BR83,DGMA85,KR90,SKM93,NB00,Cyburt04},  stellar explosions~\citep{HSSMS03,PJMI08,LICNUCF10,Longland12,BM12,BAMP20}, and the r-process~\citep{MMS12,MSFBMKA15,SPSMMS20,Barnes21}. Most of these contributions are based on direct Monte Carlo simulations and their focus is on understanding the impact of nuclear reaction rate and/or other nuclear uncertainties on the resulting nucleosynthesis predictions. 

The goal of this work is to develop skeletal RNs suitable for situations pertaining to SNe Ia. A skeletal model is a subset of a detailed reaction model which is generated by eliminating unimportant isotopes and reactions~(\cite{Smooke91,Peters1993,SFCFR16,LCW20}). The skeletal reduction is usually the first step in developing a model reduction. The next steps in the reduction include  time-scale analysis techniques, \textit{e.g.} quasi steady state approximation~(\cite{Stiefenhofer98,GI14}), partial equilibrium approximation~(\cite{Rein92,Goussis12}), and rate controlled constrained equilibrium~(\cite{Keck90,HJSM16}) amongst others. To develop skeletal RNs for SNe Ia, first local sensitivities of isotope mass fractions with respect to reaction rates are analyzed during the constant density and temperature (constant-$\rho T$) burning of carbon and oxygen with different initial conditions. The isotopes are ranked based on their sensitivities (importance), and several sets of skeletal models with different levels of accuracy are generated by selecting different numbers of important isotopes.  The sensitivities are computed by the forced optimally time-dependent (f-OTD) methodology \citep{DCB22}. This is an on-the-fly reduced order modeling (ROM) technique, recently introduced for computing sensitivities in evolutionary dynamical systems. Unlike the traditional ROM techniques, the f-OTD does not require any offline data generation, and all the computations are carried out online. \cite{NBGCL22} and \cite{LBGCLN2024} conducted a similar sensitivity-based skeletal kinetics reduction technique for chemical combustion which  automatically eliminates unimportant reactions and species. Time-dependent f-OTD modes are able to capture sudden transitions associated with the largest finite-time Lyapunov exponents \citep{BFHS17}.  Time-dependent bases have also been used for stochastic reduced order modeling \citep{SL09,CHZI13,Babaee:2017aa,B19,PB20} and on-the-fly reduced order modeling of reacting species transport equation \citep{RNB21,ANGB22}. The f-OTD can be formulated as a special case of the dynamical low-rank approximation~(\cite{KL07}). The specific objectives here are $i$) to introduce the f-OTD technique for computing sensitivities for a nuclear combustion system, and $ii$) to find skeletal models for thermonuclear burning in SNe Ia. The first set of skeletal models are applicable to both neutron-rich and equal numbers of neutron and proton scenarios. This is facilitated by the f-OTD skeletal reduction technique, without \textit{a priori} assumptions or expertise, \textit{e.g.} the assumption of an equal number of protons and neutrons. Section~\ref{sec:skeletal_fOTD}  briefly presents the theory behind the f-OTD method for constant-$\rho T$ burning in SNe Ia and the automatic process of eliminating unimportant isotope/reaction from a detailed RN. This elimination process is explained in Section~\ref{sec:aprox21} with a simple example, starting from a RN with 21 isotopes and reducing it to a skeletal model with 10 isotopes. Section~\ref{sec:torch} describes the application of the f-OTD skeletal reduction method to the Torch RN\footnote{\url{https://cococubed.com/code_pages/net_torch.shtml}}  \citep{Timmes99,TS00,AHGLF19}. This RN considers 495 isotopes, up to $^{91}\text{Tc}$, and 6012 reactions. Different skeletal models with different levels of accuracy are extracted and their ability to reproduce the energy release and $^{44}\text{Ti}$ and $^{56}\text{Ni}$ abundances of the Torch model are analyzed. Section~\ref{sec:conclusion} provides the concluding remarks.

\section{Skeletal reduction with f-OTD method} \label{sec:skeletal_fOTD}

\subsection{Reduced-order modeling of the sensitivity matrix with f-OTD}
\label{subsec:fOTD_modeling}

For the model description, let isotope $k$ have total charge $Z_k$ and atomic weight $A_k$. Let the aggregate total of isotope $k$ have a mass density $\rho_k$ and a number density $\mathtt{n}_k$ in a material with the temperature $T$ and the total mass density $\rho$.  The mass fraction of isotope $k$ is defined as $\mathtt{x}_k= \rho_k / \rho = \mathtt{n}_k A_k/\rho N_A$ where $N_A = 6.02252 \times 10^{23}$ particles/mol is the Avogadro's number. The mean atomic weight is $A_m= (\sum \mathtt{x}_k / A_k)^{-1}$, and is the equivalent of the mixture molar
mass from the combustion literature \citep{Williams85}. The nuclear abundance of isotope $k$ is $\mathtt{y}_k = \mathtt{x}_k/A_k = \mathtt{n}_k/(\rho N_A)$, the mean charge per isotope is $Z_m= A_m\sum Z_k \mathtt{x}_k / A_k$, and the electron abundance, or electron number fraction, is $\mathtt{y}_{e} = Z_m /A_m$. This is related to the neutron excess, $\eta$, by $\eta=1-2\mathtt{y}_{e}$, so that $\eta=0$ corresponds to $\mathtt{y}_{e}=0.5$. The total scalar pressure, the total specific internal energy, and the total specific entropy are denoted by $p_{tot}$, $e_{tot}$, and $s_{tot}$, respectively \citep{NGL19}. Other quantities such as the specific heats or adiabatic indices can be determined via an equation of state \citep{TA99} once the partial derivatives of the pressure and the specific internal energy with respect to the density and temperature are known. Consider a nuclear system of $n_s$ isotopes reacting through $n_r$  reactions:
\begin{equation}
\sum_{k=1}^{n_s} \nu'_{kj} \mathbb{M}_{k} \leftrightharpoons
\sum_{k=1}^{n_s} \nu''_{kj} \mathbb{M}_{k}, \ \ j=1, \dots n_r,
\end{equation}

\noindent where $\mathbb{M}_{k}$ is a symbol for isotope $k$, and $\nu'_{kj}$ and $\nu''_{kj}$ are the stoichiometric coefficients of isotope $k$ in reaction $j$. Changes of abundances $\bxi = [\mathtt{y}_1, \mathtt{y}_2, \dots, \mathtt{y}_{n_s}]^T$ in constant-$\rho T$ burning within a carbon-oxygen white dwarf (WD) with constant temperature and pressure can be described by the following initial value problem \citep{THW00}:
\begin{equation}
\label{eq:xi_IVP}
 \frac{d\mathtt{y}_k}{dt} = f_{k}(\bxi,\balpha)  = \frac{A_k}{N_A} \sum_{j=1}^{n_r} \nu_{kj} \alpha_j \mathcal{Q}_j, \  \  \bxi(0) = \bxi_0,
\end{equation}
\noindent where $t \in [0, t_{f}]$ is time, $t_f$ is the final time, and $\balpha = [1,1, \dots, 1] \in \mathbb{R}^{n_r}$ is the vector of perturbation parameters. In Eq.\ (\ref{eq:xi_IVP}), $\mathcal{Q}_j$ is the progress rate of reaction $j$ and is equal to the following for one and two body reactions:
\begin{subequations}
\label{eq:Q_nu}
\begin{eqnarray} 
 \nu_{kj} &=& \nu''_{kj} - \nu'_{kj}, \\
 \mathcal{Q}_j&=&K_{f,j} \prod_{k=1}^{n_s} \lp \mathtt{n}_k \rp^{\nu'_{kj}}-K_{r,j} \prod_{k=1}^{n_s} \lp \mathtt{n}_k \rp^{\nu''_{kj}} \non \\&=&K_{f,j} \prod_{k=1}^{n_s} \lp \rho \mathtt{y}_k N_A \rp^{\nu'_{kj}}-K_{r,j} \prod_{k=1}^{n_s} \lp \rho \mathtt{y}_k N_A \rp^{\nu''_{kj}}.
\end{eqnarray}
\end{subequations}
Here, $K_{f,j}$ and $K_{r,j}$ are the forward and reverse rate of reaction $j$. The quantity $\prod_{k=1}^{n_s} \lp \mathtt{n}_k \rp^{\nu'_{kj}}$ is the total possible number of elementary reactions per unit volume (obtained by counting the number of possible collisions) and is based on the number density of particles. Because the collision energies are well below the Coulomb barrier, most collisions do not result in nuclear reactions. Thus, the reaction rate is the product of the collision rate and the tunneling probability.

The abundances in Eq.\ (\ref{eq:xi_IVP}) are perturbed by infinitesimal variations of $\alpha_j$, by letting $\alpha_j=1+\delta \alpha_j$, where $\delta \alpha_j \ll 1$ for $j=1,2,\dots, n_r$. The perturbation with respect to $\alpha_j$ amounts to an infinitesimal perturbation of progress rates $\mathcal{Q}_j$.  The sensitivity matrix, $S(t)=[\bs_1(t), \bs_2(t), ... \bs_{n_r}(t)]\in \mathbb{R}^{n_s \times n_r}$, contains local sensitivity coefficients, $\bs_{j}=\prt \bxi / \prt \alpha_j$, and can  be calculated by solving the sensitivity equation (SE),
\begin{equation}
 \label{eq:sensit_exact}
 \begin{split}
\frac{dS_{ij}}{dt} &=  \sum_{m=1}^{n_s} \frac{\prt f_{i}}{\prt \mathtt{y}_{m}} \frac{\prt \mathtt{y}_{m}}{\prt \alpha_{j}} + \frac{\prt f_{i}}{\prt \alpha_{j}}  = \sum_{m=1}^{n_s} L_{im} S_{mj} + F_{ij}, 
\end{split}
\end{equation}
\noindent where $L_{im} =  \frac{\prt f_{i}}{\prt \mathtt{y}_{m}}$ and $F_{ij}=\frac{\prt f_{i}}{\prt \alpha_{j}}$ are the Jacobian and the forcing matrices, respectively.
\begin{figure}[!htbp]
\centering
 \includegraphics[width=14cm]{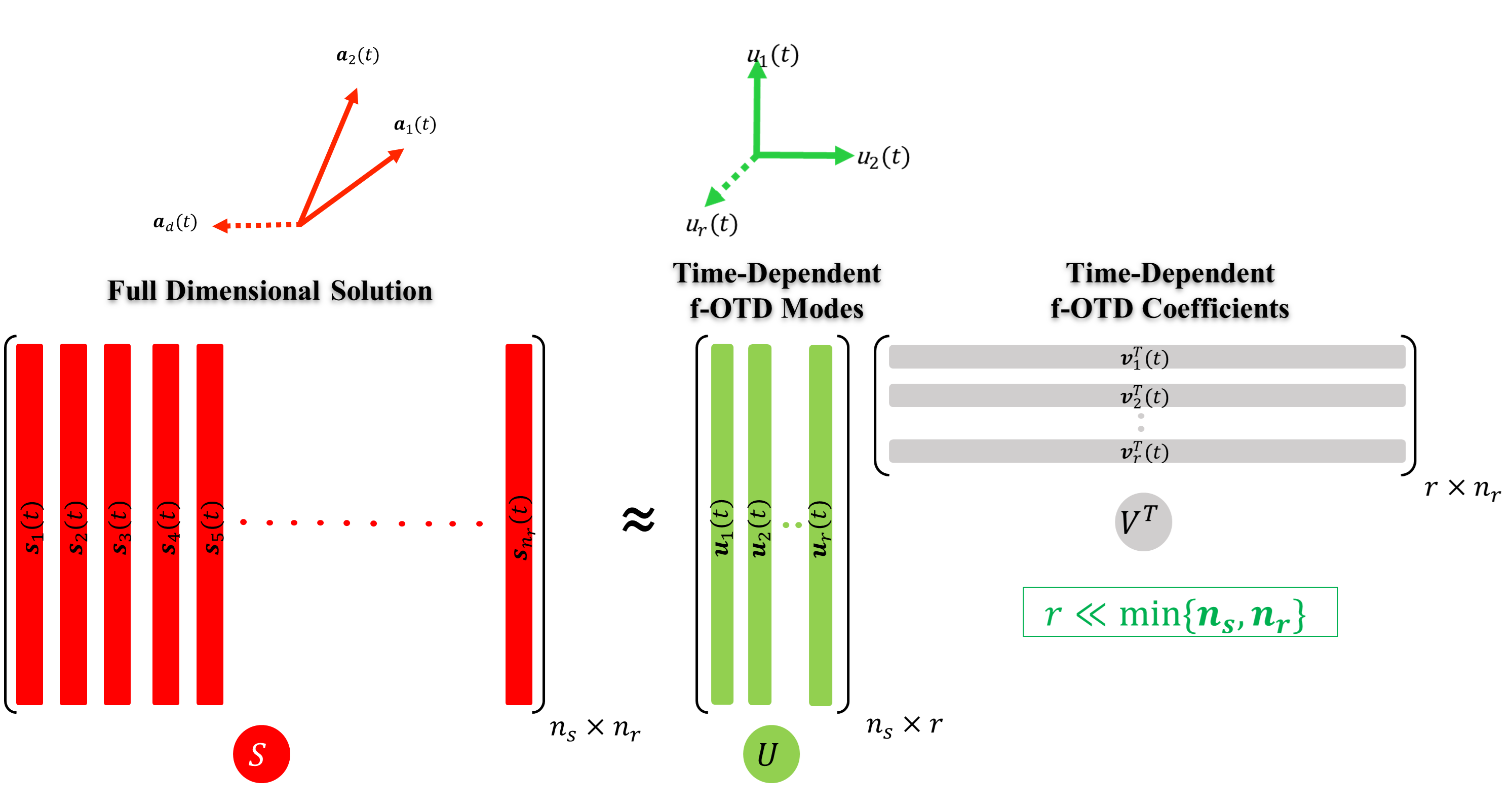}
 \caption{Modeling sensitivity matrix $S(t)$ as a multiplication of two low-ranked matrices $U(t)$ and $Y(t)$ which evolve according Eq.\ (\ref{eq:UY_evolution}). Reprinted from \cite{NBGCL22} with permission.}
\label{FIG:S_matrix_shape}
\end{figure}
The model reduction strategy is based on selecting reactions, whose perturbations grow  most intensely in the abundance Eq.\ (\ref{eq:xi_IVP}). The selection of important reactions is performed by \textit{instantaneous} observation of \textit{modeled} sensitivities. In f-OTD, the sensitivity matrix $S(t)$ is modeled by factorizing it into a time-dependent subspace in the $n_s$-dimensional phase space of abundances represented by a set of f-OTD modes: $U(t) = [\bu_1(t), \bu_2(t), \dots, \bu_r(t)] \in \mathbb{R}^{n_s \times r}$. These modes are orthonormal $\bu_i^T(t) \bu_j(t) = \delta_{ij}$ at all $t$, where $\delta_{ij}$ is the Kronecker delta. The rank of $S(t)\in \mathbb{R}^{n_s \times n_r}$ is $d=\mbox{min}\{n_s,n_r\}$ while the f-OTD modes represent a rank-$r$ subspace, where $r\ll d$. To this end,  the sensitivity matrix is approximated via the f-OTD decomposition (Fig.\ \ref{FIG:S_matrix_shape}) as $S(t) \approx U(t)V^T(t)$ where $V(t) = [\bvv_1(t), \bvv_2(t), \dots, \bvv_r(t)] \in \mathbb{R}^{n_r \times r}$ is the f-OTD coefficient matrix. This decomposition is a low-rank approximation of the sensitivity matrix $S(t)$. Therefore, $U(t)V(t)^T$ closely approximates $S(t)$  and it is not exact.  Both $U(t)$ and $V(t)$ are time-dependent, and their explicit time dependency on $t$ is dropped for brevity. Figure~\ref{FIG:S_matrix_shape} shows the schematic of the decomposition of $S$ into f-OTD components $U$ and $V$. The evolution equation for $U$ and $V$ are obtained by substituting the sensitivity decomposition ($S(t) \approx U(t)V^T(t)$) into Eq.\ (\ref{eq:sensit_exact}):
\begin{subequations}
\label{eq:UY_evolution}
\begin{eqnarray} 
\label{eq:U_evolution}
\frac{dU}{dt} &=& QLU+QFVC^{-1}, \\
\label{eq:Y_evolution}
\frac{dV}{dt} &=& V L_r^T +  F^T U,
\end{eqnarray}
\end{subequations}
\noindent where $Q= I-UU^{T}$ is the orthogonal projection onto the space spanned by the complement of $U$ and $C=V^TV \in \mathbb{R}^{r \times r}$ is a correlation matrix. Matrix $C(t)$ is, in general, a full matrix implying that the f-OTD coefficients are correlated. $L_r = U^T L U \in \mathbb{R}^{r \times r}$ is a reduced linearized operator. Equation (\ref{eq:UY_evolution}) represents a coupled system of ODEs and constitutes the f-OTD evolution equations. The f-OTD modes align themselves with the most instantaneously sensitive directions of the abundance evolution equation when perturbed by $\balpha$. It is shown by \cite{BFHS17} that when $\balpha$ is the perturbation to the initial condition, the OTD modes converge exponentially to the eigen-directions of the Cauchy–Green tensor associated with the most intense finite-time instabilities.

The primary computational advantage of using f-OTD is that the method only evolves two skinny matrices containing $(n_s+n_r)\times r$ elements as opposed to $n_s \times n_r$ elements in the SE (Eq.\ (\ref{eq:sensit_exact})). This reduces the required memory for ODE solvers drastically and facilitates the application of stiff  solvers for evolving sensitivities. Moreover, in the f-OTD decomposition, the sensitivities are stored in the \emph{compressed form}, \textit{i.e.}, matrices $U$, and $V$ are kept in the memory as opposed to their multiplication $U V^T$, \textit{i.e.}, the \emph{decompressed form}. Therefore,  in comparison to the full SE, f-OTD decomposition results in the memory compression ratio of $(n_s \times n_r)/((n_s+n_r)r)$.

\subsection{Identification of important reactions \& isotopes}
\label{subsec:fOTD_reac_selec}

In f-OTD skeletal reduction, modeled sensitivities are computed in a factorized format by solving Eqs.\ (\ref{eq:xi_IVP}), (\ref{eq:U_evolution}), and (\ref{eq:Y_evolution}), and the values of $U$, $V$, and $\bxi$ are stored at resolved time steps $t_i \in [0,t_f]$. Equation\  (\ref{eq:xi_IVP}) is initialized with different sets of isotope abundances, temperature and density within their ranges of interest. Each simulation with a different initial condition is identified as a \textit{case}. At each resolved time step and for each case, the eigen decomposition of $S^T S\in \mathbb{R}^{n_r \times n_r}$ is computed as $A \Lambda A^T$, and the vector $\bw  = ( \Sigma \lambda_i |\baa_i|)/( \Sigma \lambda_i) \in \mathbb{R}^{n_r}$  is basically the average of eigenvectors of $S^T S$ matrix weighted based on their associated eigenvalues ($\lambda_i$), and prevents dealing with each eigenvector ($\baa_i$) separately. Each component of $\bw$, \textit{i.e.} $w_i$, is positive and associated with a certain reaction ($i$th reaction). The larger the $w_i$ value, the more important the reaction $i$ is. The $w_{max,i}$ denotes the highest value of $w_i$ through all resolved time steps and cases. The elements of $\bw_{max}$ vector are sorted in descending order to find the indices of the most important reactions in the detailed model. Isotopes are also sorted based on their first presence in the sorted reactions, \textit{i.e.} isotopes which first show up  in a higher ranked reaction would be more important than an isotope which first participates in a lower ranked reaction. This results  in a reaction and isotope ranking based on $\bw_{max}$ vector. Finally, a set of isotopes are chosen by defining a threshold $\epsilon$ on the element of $\bw_{max}$ vector and eliminating isotopes whose associated $w_{max,i}$ are less than $\epsilon$. This terminates reactions which include the eliminated isotopes from the detailed model. Since the model reduction is reaction based, any non-reactive isotope with a non-zero mass fraction in the initial condition must be manually added to the skeletal model.

In summary, the nuclear combustion system is instantaneously observed, and the reactions are sorted based on their effects on sensitivities to find the most important isotopes. These isotopes and the reactions which connect them together create the skeletal models.  In combustion systems, perturbations with respect to ``fast" reactions generate very large sensitivities for short time periods which vanish as $t\rightarrow\infty$. On the other hand, perturbations with respect to ``slow" reactions generate smaller and more sustained sensitivities. The approach here is based on the instantaneous observation of sensitivities, both slow and fast reactions can leave an imprint on the instantaneous normalized reaction vector ($\bw$) if their imprints are larger than the threshold value ($\epsilon$). However, if the sensitivities associated with fast and slow reactions from pre-determined times and locations are combined with each other before dimension reduction, as commonly done in principal component analysis (PCA) type schemes, the smaller sensitivities associated with slow reactions would be out-weighted by the large sensitivities associated with fast reactions.

\section{skeletal reduction on the Approx21 RN}\label{sec:aprox21}

The process of eliminating unimportant reactions and isotopes from a kinetics model with f-OTD is demonstrated in this section with a simple example. Let us start with the Approx21 model which is the default MESA network \citep{MESA,Paxton15} for alpha chain reactions. Approx21 evolves $n_s=21$ isotopes: $\ce{n}$, $\ce{p}$, $\ce{^{1}H}$, $\ce{^{3}He}$, $\ce{^{4}He}$, $\ce{^{12}C}$, $\ce{^{14}N}$, $\ce{^{16}O}$, $\ce{^{20}Ne}$, $\ce{^{24}Mg}$, $\ce{^{28}Si}$, $\ce{^{32}S}$, $\ce{^{36}Ar}$, $\ce{^{40}Ca}$, $\ce{^{44}Ti}$, $\ce{^{48}Cr}$, $\ce{^{56}Cr}$, $\ce{^{52}Fe}$, $\ce{^{54}Fe}$, $\ce{^{56}Fe}$, and $\ce{^{56}Ni}$ through $n_r=112$ reactions. In this RN, ($\alpha$,p)(p,$\gamma$) and ($\gamma$,p)(p,$\alpha$) links are included in order to obtain reasonably accurate energy generation rates and abundance levels when the temperature exceeds 2.5e9 K. At these elevated temperatures, the flows through the ($\alpha$,p)(p,$\gamma$) sequences are faster than the flows through the ($\alpha$,$\gamma$) channels. An ($\alpha$,p)(p,$\gamma$) sequence is, effectively, an ($\alpha$,$\gamma$) reaction through an intermediate isotope. By assuming steady-state proton flows through intermediate isotopes $\ce{^{27}Al}$, $\ce{^{31}P}$, $\ce{^{35}Cl}$, $\ce{^{39}K}$, $\ce{^{43}Sc}$, $\ce{^{47}V}$, $\ce{^{51}Mn}$, and $\ce{^{55}Co}$, this strategy avoids explicitly evolving the abundances of the proton or intermediate isotopes\footnote{\url{https://cococubed.com/code_pages/burn_helium.shtml}}. The skeletal reduction is exemplified here by analyzing only one case for the constant-$\rho T$ burning of a mixture of carbon and oxygen in a WD progenitor. 
\begin{figure}[!h]
%\begin{wrapfigure}{r}{9cm}
\centering
 \epsfig{file=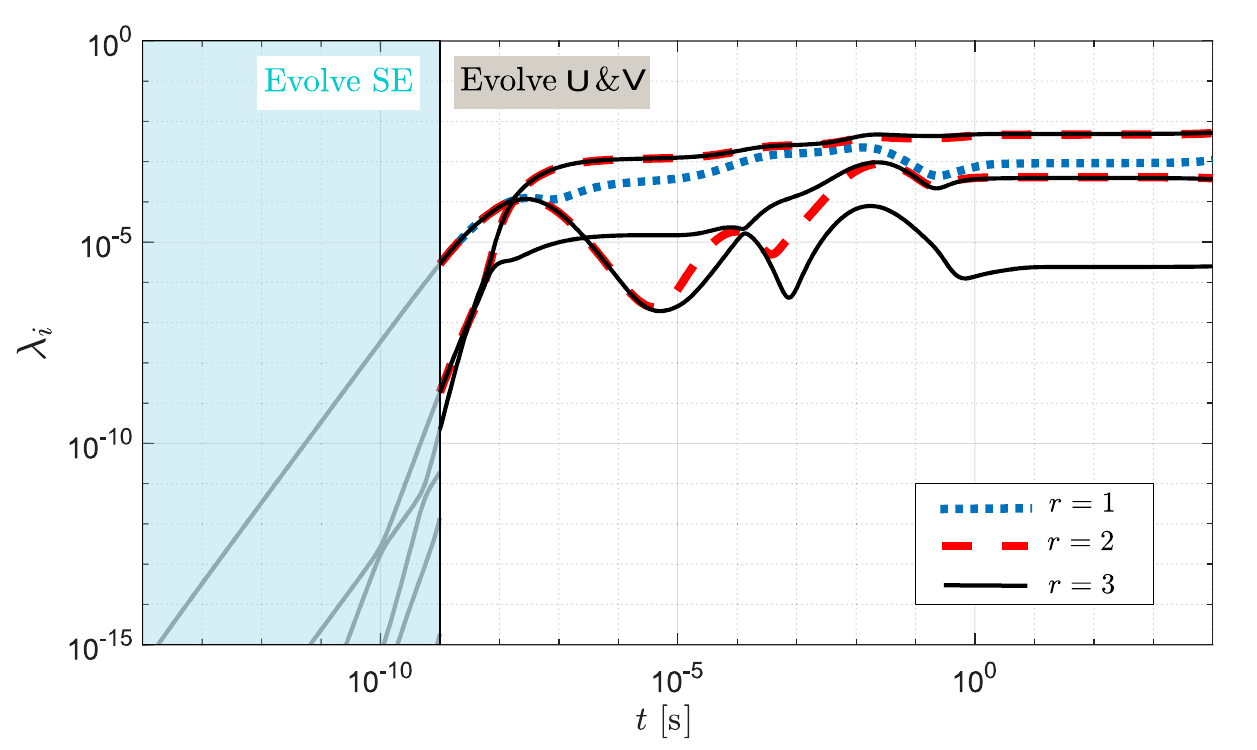,width=6cm,angle =-90}
 \caption{Model reduction for Approx21: eigenvalues of the $S^TS$ matrix. The sensitivity matrix, $S$, initially evolves exactly via the SE (Eq.\ (\ref{eq:sensit_exact})), and then evolves approximately with the f-OTD equations (Eq.\ (\ref{eq:UY_evolution})). The f-OTD simulation ($t>10^{-9}$) only evolves $r$ modes of the full order model ($t<10^{-9}$). Ignition data is gathered from the constant-$\rho T$ burning  simulation case described in Fig.\ \ref{fig:Aprox21_ranking}.} 
\label{fig:Sigm_aprox21}
\end{figure}
%\end{wrapfigure}
The ignition initiates with $T_9=3$, $\rho_9=1$, and $\mathtt{x}_{C,0}=\mathtt{x}_{O,0}=0.5$, so $\mathtt{y}_{e}=0.5$. Here, $T_9\equiv T/(10^9 K)$ and $\rho_9 \equiv \rho/(10^9 g.cm^{-3})$. The sensitivities evolve in two phases. In the first, the full-dimensional SE (Eq.\ (\ref{eq:sensit_exact})) is solved for a very small time period, \textit{e.g.} until $t=10^{-12} s$, to generate the initial conditions for the f-OTD simulation. In the second phase, the sensitivity matrix ($S$) is approximated  by evolving the f-OTD equations (Eq.\ (\ref{eq:UY_evolution})). The $U$ and $Y$ matrices are initialized by eigenvalue decomposition of the full sensitivity matrix ($S$) at the end of the first phase. Figure\ \ref{fig:Sigm_aprox21} shows the evolution of the eigenvalues of $S^TS$ matrix. It is indicated that $i$) $\lambda_1$ is an order of magnitude larger than $\lambda_2$ most of the time, and $ii$) the modeled sensitivities converge by adding more modes. Therefore, f-OTD simulation with $r=1$ provides a reasonable estimation of sensitivities, which is enough if the final goal is to determine the importance of reactions/isotopes. 

\begin{figure}[!h]
\centering
 \epsfig{file=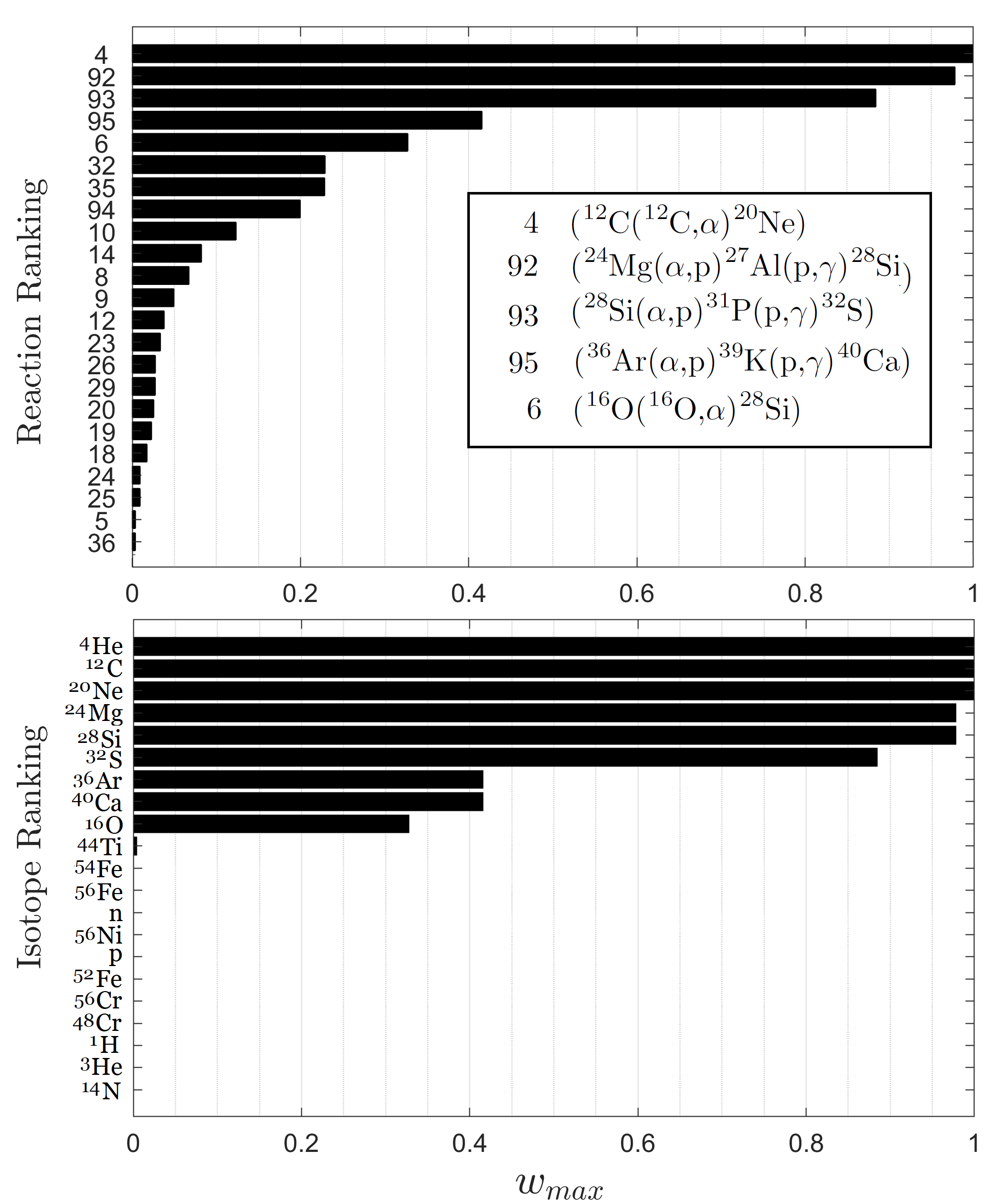, width=8.0cm}
 \caption{Model reduction for Approx21: reaction and isotope ranking based on their associated $w_{max}$ from one constant-$\rho T$ burning simulation case with $T_{9}=3$, $\rho_9=3$, and initial composition of $\mathtt{x}_{C,0}=\mathtt{x}_{O,0}=0.5$ with $\mathtt{y}_{e,0}=0.5$.}
\label{fig:Aprox21_ranking}
\end{figure}

\begin{figure}[!h]
\centering
 \epsfig{file=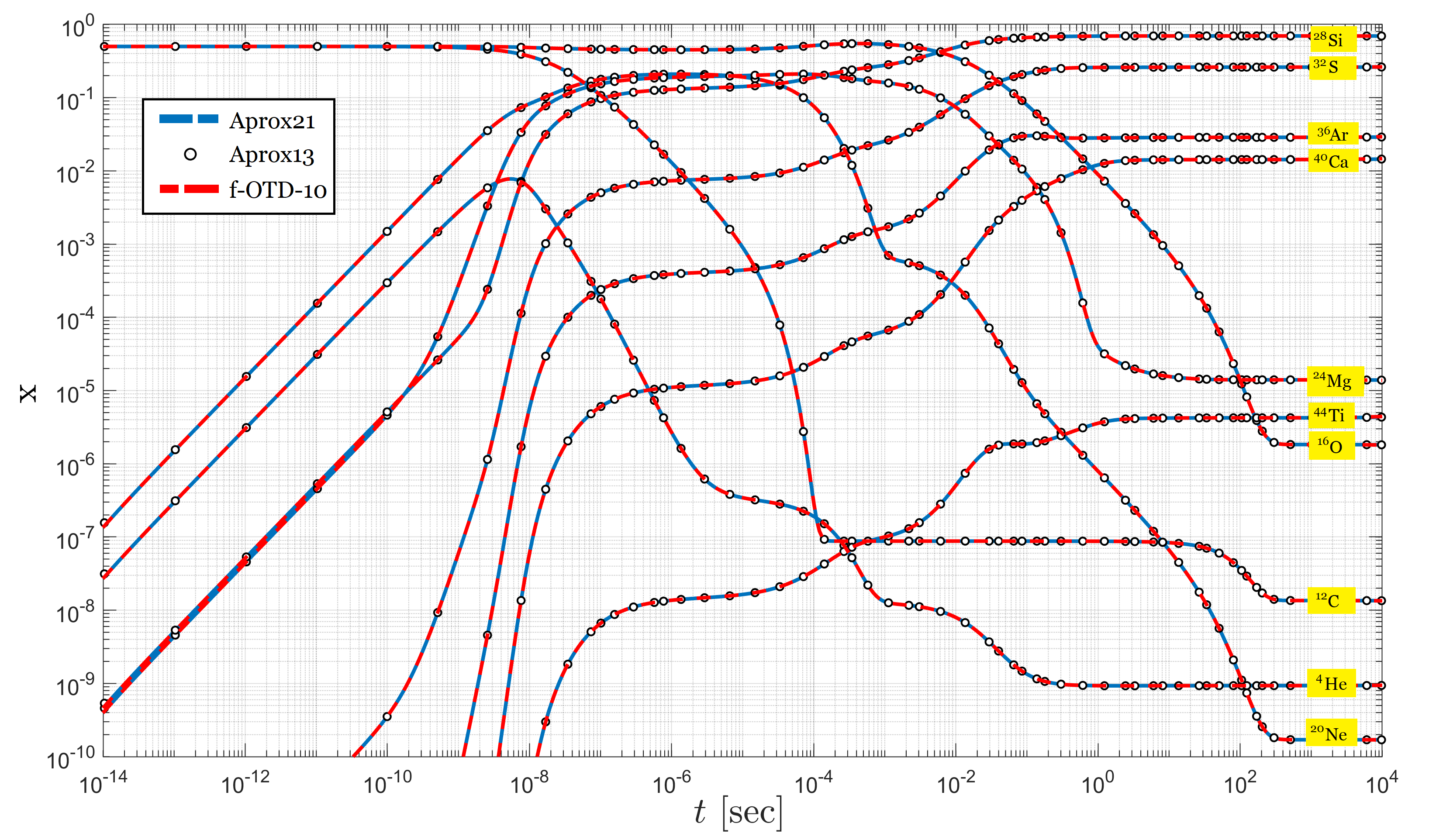,width=12.0cm}
 \caption{Model reduction for Approx21: evolution of isotope mass fractions based on Approx21, Approx13, and f-OTD-10. The last two models are generated from Approx21.}
\label{fig:Aprox21_Psi}
\end{figure}

Figure\ \ref{fig:Aprox21_ranking} shows the ranking of reactions and isotopes in Approx21 associated with the constant-$\rho  T$ burning case. It is apparent that reactions  4 ($\ce{^{12}C}$($\ce{^{12}C}$,$\alpha$)$\ce{^{20}Ne}$), equilibrium reactions 92 ($\ce{^{24}Mg}$($\alpha$,p)$\ce{^{27}Al}$(p,$\gamma$)$\ce{^{28}Si}$), 93 ($\ce{^{28}Si}$($\alpha$,p)$\ce{^{31}P}$(p,$\gamma$)$\ce{^{32}S}$), and 95 ($\ce{^{36}Ar}$($\alpha$,p)$\ce{^{39}K}$(p,$\gamma$)$\ce{^{40}Ca}$), and reaction 6 ($\ce{^{16}O}$($\ce{^{16}O}$,$\alpha$)$\ce{^{28}Si}$) are the first five most important reactions, and isotopes  $\ce{^{4}He}$, $\ce{^{12}C}$, $\ce{^{20}Ne}$, $\ce{^{24}Mg}$, and $\ce{^{28}Si}$  are the first five most important isotopes in Approx21 for the ignition case considered.
%\begin{wrapfigure}{r}{9cm}
%\end{wrapfigure}
The f-OTD-10 model is then created based on the algorithm described in \S\ref{subsec:fOTD_reac_selec} and contains 10 most important isotopes shown in Fig.\ \ref{fig:Aprox21_ranking}, \textit{i.e.},  $\ce{^{4}He}$, $\ce{^{12}C}$,  $\ce{^{20}Ne}$, $\ce{^{24}Mg}$, $\ce{^{28}Si}$, $\ce{^{32}S}$, $\ce{^{36}Ar}$, $\ce{^{40}Ca}$, $\ce{^{16}O}$, and $\ce{^{44}Ti}$. Approx13 (\cite{THW00}) is a 13 isotope ($\ce{^{4}He}$, $\ce{^{12}C}$, $\ce{^{16}O}$, $\ce{^{20}Ne}$, $\ce{^{23}Mg}$, $\ce{^{28}Si}$, $\ce{^{32}S}$, $\ce{^{36}Ar}$, $\ce{^{40}Ca}$, $\ce{^{44}Ti}$, $\ce{^{48}Cr}$, $\ce{^{52}Fe}$, and $\ce{^{56}Ni}$) reaction network which is also extracted from Approx21 but over a wider time range to produce $\ce{^{56}Ni}$. That is why Approx13 contains three more isotopes, \textit{i.e.} $\ce{^{48}Cr}$, $\ce{^{52}Fe}$, and $\ce{^{56}Ni}$ in comparison with f-OTD-10. Nevertheless, this simple exercise demonstrates the ability of f-OTD methodology to extract the relevant isotopes for a given set of conditions and starting RN. Figure~\ref{fig:Aprox21_Psi} demonstrates the performance of f-OTD-10 RN in predicting the evolution of isotope mass fractions over its design conditions ($T_9=3$, $\rho_9 =1$, $\mathtt{x}_{C,0}=\mathtt{x}_{O,0}=0.5$, and $t \in [0,10^4]s$). 

\section{Skeletal reduction on the Torch RN}\label{sec:torch}

In SNe Ia, the  carbon and oxygen burn together to produce nuclei from silicon to the iron peak which are being ejected into the interstellar medium \citep{Nomoto97,WHW02,LR17,Johnson19}. As this type of supernova does not produce a significant amount of free neutrons, it does not synthesize elements beyond the iron peak \citep{Johnson19}. Nevertheless, some heavier isotopes can contribute to reactions involving isotopes important to light curves observations, such as $\ce{^{56}Ni}$. The 495-isotope version of the Torch RN \citep{Paxton15} is chosen to represent the detailed RN in this section. The Torch RN extends from  $\ce{^{1}H}$ to $\ce{^{91}Tc}$ \citep{Timmes99} with $n_s=495$ and $n_r=6012$. The weak reactions are turned on, and no screening is performed on the reaction rates \citep{FFN85}. The Helmholtz EOS as developed by \cite{TS00} is used with Coulomb correction to calculate the internal energy and the pressure. Several skeletal RNs have been previously proposed based on the Torch model for inline calculations \citep{Timmes99,TS00,AHGLF19} and are used in the MESA code \citep{MESA,AHGLF19,Paxton15}. These include a bare minimum model of the $\alpha$-chain reactions using 13 isotopes, a 19-isotope RN to also accommodate some hydrogen burning \citep{Weaver1978}, and a 21-isotope RN that adds $^{56}$Cr and $^{56}$Fe and respective equilibrium reaction sequences to the 19 isotope network to attain a lower $\mathtt{y}_e$ value for pre-supernova models \citep{Paxton15}. Several important isotopes are produced in burning scenarios with $\mathtt{y}_{e}$ significantly lower than $0.5$. For example, \cite{Woosley97} suggests that $^{48}$Ca can only be produced in nature in a subset of SNe Ia, with $\mathtt{y}_{e}$ in the range $0.41$ to $0.42$ and high burning density. The Torch RN covers such scenarios. On the other hand, the performances of the existing skeletal models have not been systematically examined to cover both $\mathtt{y}_{e}=0.5$ and $\mathtt{y}_{e}<0.5$ scenarios. This is addressed here by considering initial conditions with different $\mathtt{y}_{e}$ values and choosing the 21-isotope RN skeletal model (hereafter denoted Approx21) for comparison with the proposed skeletal RNs. 

\begin{figure}[!h]
\centering
 \epsfig{file=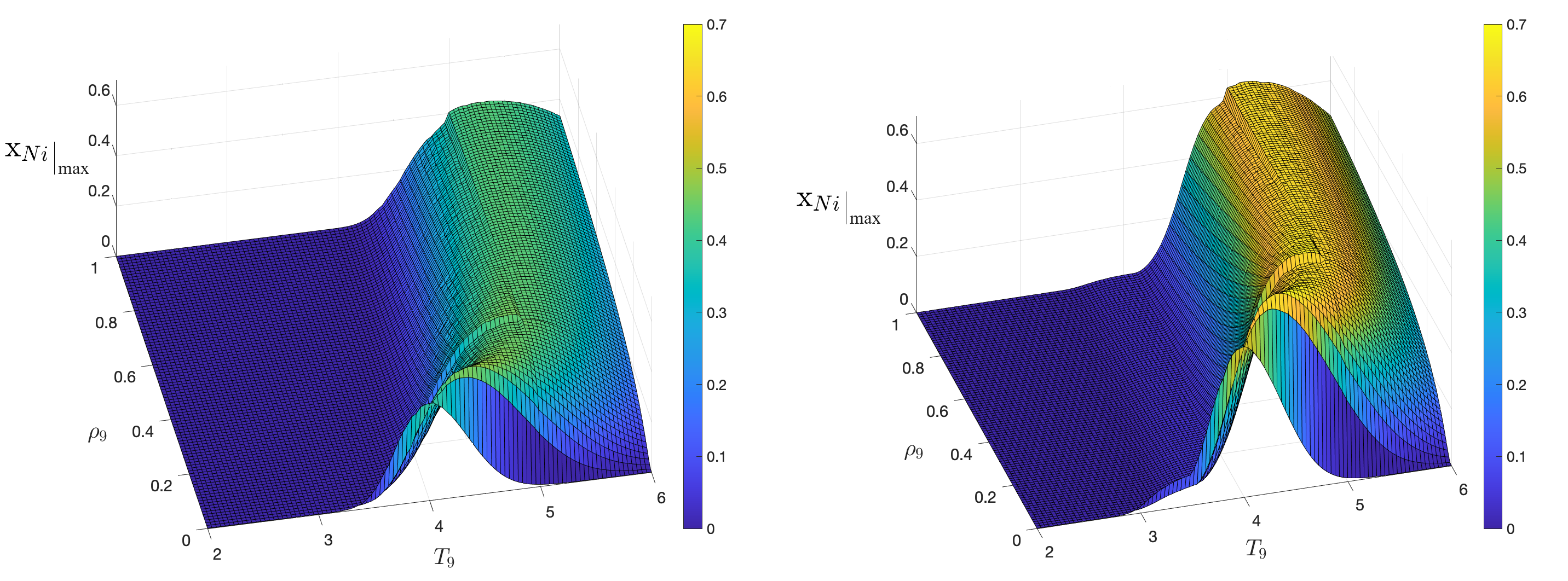,width=16.0cm}
 \caption{Maximum mass fraction of $^{56}\text{Ni}$ during constant-$\rho T$ burning of SNe Ia. Red circles show the f-OTD cases.}
\label{fig:Torch_Ni_max}
\end{figure}

\begin{figure}[!h]
\centering
 \epsfig{file=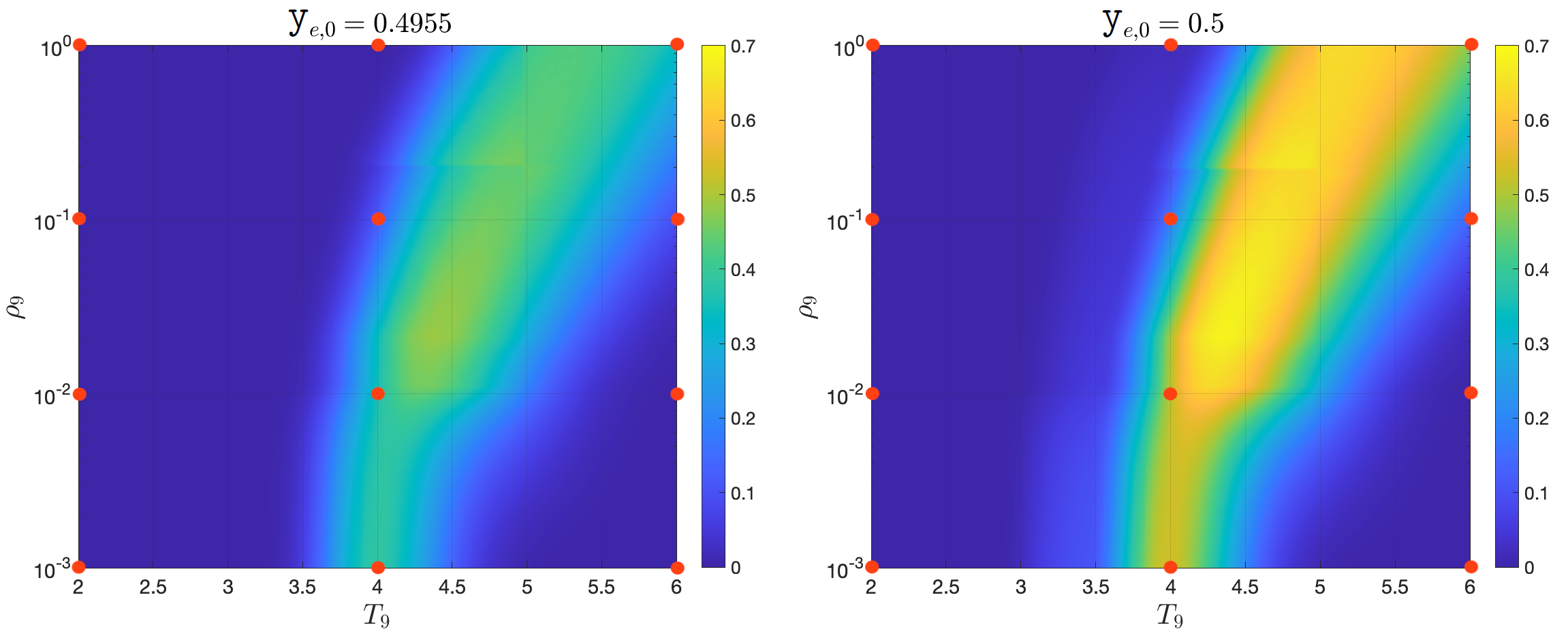,width=16.0cm}
 \caption{Maximum mass fraction of $^{56}\text{Ni}$ during constant-$\rho T$ burning of SNe Ia. Red circles show the f-OTD cases.}
\label{fig:Torch_Ni_max_2D}
\end{figure}

The SNe Ia progenitor population and burning scenarios cover a wide range of temperatures and densities \citep{HN00} so that, most likely, a single skeletal model with a limited number of isotopes ($n_s$) cannot yield accurate predictions over the full range of conditions which would be covered by a detailed RN. Because of the importance of $^{56}\text{Ni}$ in SNe Ia, the skeletal models in this work are designed to predict the evolution of $\mathtt{x}_{^{56}\text{Ni}}$ correctly. For this purpose, a map of maximum production of this isotope is produced during the course of constant-$\rho T$ burning in SNe Ia as shown in Fig.\ \ref{fig:Torch_Ni_max}. The Torch RN is run 10000 times on a 100$\times$100 grid of $T_9$ and $\rho_9$ for  $\mathtt{y}_{e,0}$ values of 0.4955 and 0.5. It is observed that $^{56}\text{Ni}$ is only significantly produced within certain ranges of the temperature and the density values, with a noticeable shift of the $^{56}\text{Ni}$ production near the peak. In particular, the peak occurs around $T_9=4.0$ at lower densities. Moreover, the maximum production of $^{56}\text{Ni}$ is decreased by decreasing $\mathtt{y}_{e,0}$ \citep{Iliadis15}. To generate skeletal reactions, simulations are conducted of constant-$\rho T$ burning in SNe Ia for 24  cases, with $T_9 \in \{2,4,6\}, \rho_9 \in \{0.001,0.01,0.1,1.0\}$, and $\mathtt{y}_{e,0} \in \{0.4955,0.5\}$. To better show how these cases are distributed, a 2D version of Fig.\ \ref{fig:Torch_Ni_max} is shown in Fig.\ \ref{fig:Torch_Ni_max_2D}. Each case (shown as a red circle in Fig.\ \ref{fig:Torch_Ni_max_2D}) denotes a set of initial condition for the density, the temperature, the energy and isotope mass fractions.  These conditions cover the burning stage, and are of interest in multidimensional simulations \citep{FORTZLMRTT00,Woosley_etal07}.

\begin{figure}[!h]
\centering
 \epsfig{file=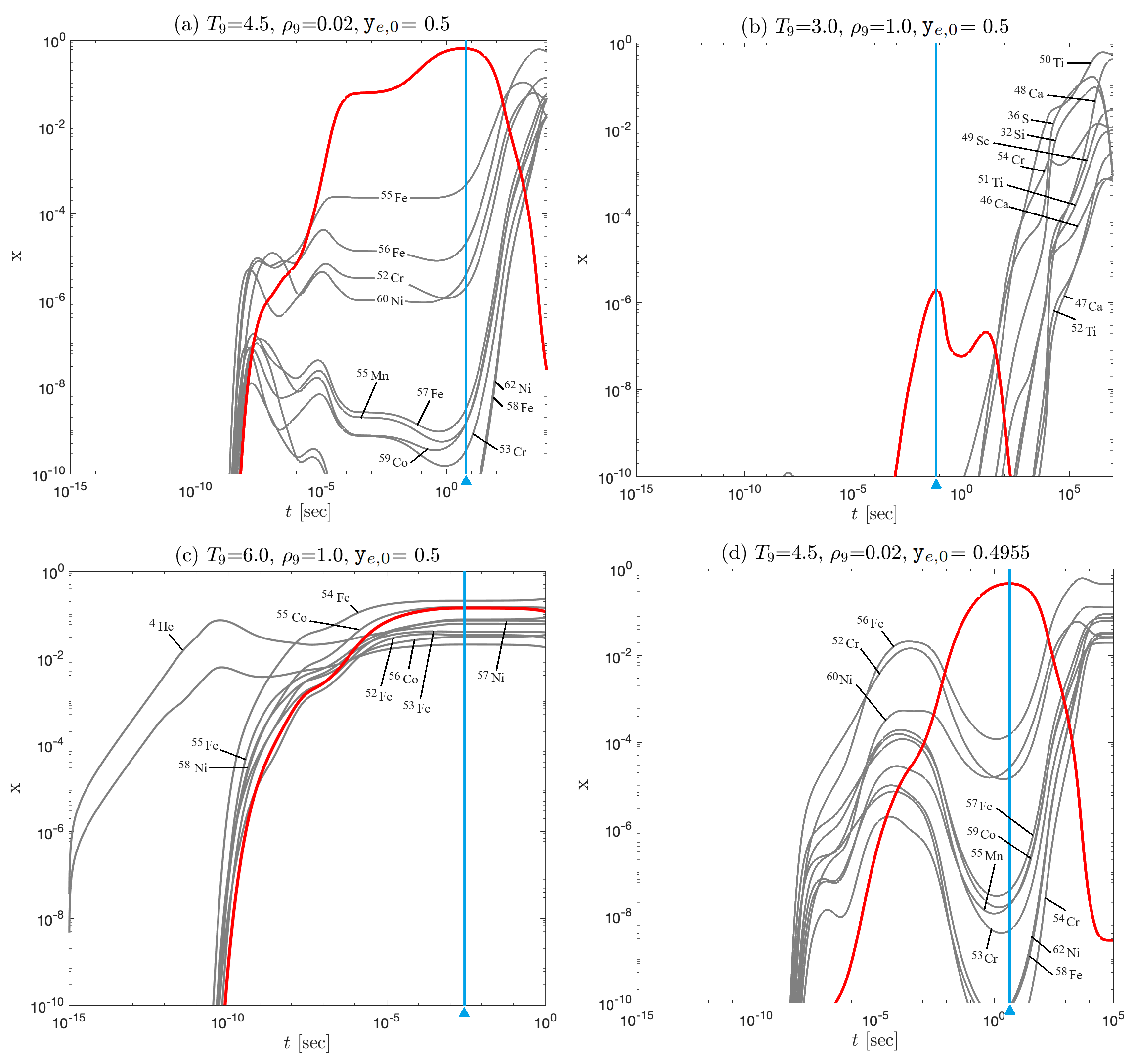,width=16.0cm}
 \caption{Evolution of isotope mass fractions in Torch with different initial conditions, highlighting different scenarios for $\ce{^{56}Ni}$ evolution. The blue lines indicate the final time of the f-OTD simulations and sensitivity analysis. The red and gray lines represent the $\ce{^{56}Ni}$ mass fraction and the 10 isotopes with the highest final mass fractions observed during the portrayed ignition process. These isotopes are listed in the left-top corner of each sub-figure }
\label{fig:t_final}
\end{figure}

 The final time for each case is when the mass fraction of $^{56}\text{Ni}$ reaches its maximum. Figure\ \ref{fig:t_final} shows the evolution of isotope mass fractions for two different cases. The red lines in Fig.\ \ref{fig:t_final} show the  $^{56}\text{Ni}$ mass fraction and blue triangles denote the final time of f-OTD simulations and sensitivity analysis. The initial mass fractions of $\ce{^{12}C}$, $\ce{^{16}O}$, and $\ce{^{22}Ne}$ isotopes in cases with $\mathtt{y}_{e,0}=0.5$ are $\mathtt{x}_{C,0}=\mathtt{x}_{O,0}=0.5$, and in cases with $\mathtt{y}_{e,0}=0.4955$ are $\mathtt{x}_{C,0}=0.45$, $\mathtt{x}_{O,0}=0.45$, and $\mathtt{x}_{Ne,0}=0.1$. Note that for the conditions considered here, the sharp decline in $^{56}\text{Ni}$ mass fraction at $t \gtrsim 10s$ is likely due to electron capture, causing the network composition to shift to more neutron-rich isotopes of nickel and iron, which is not really relevant to carbon/oxygen burning and therefore not used in the f-OTD analysis. The f-OTD method is used to model the sensitivity matrix with $r=1$ mode, and the generated skeletal models by sensitivity analysis are denoted by f-OTD-n in which ``n" identifies for the number of isotopes. The predictive capabilities of f-OTD models are compared against those obtained via Torch RN and the Approx21. 

\begin{figure}[!h]
\centering
 \epsfig{file=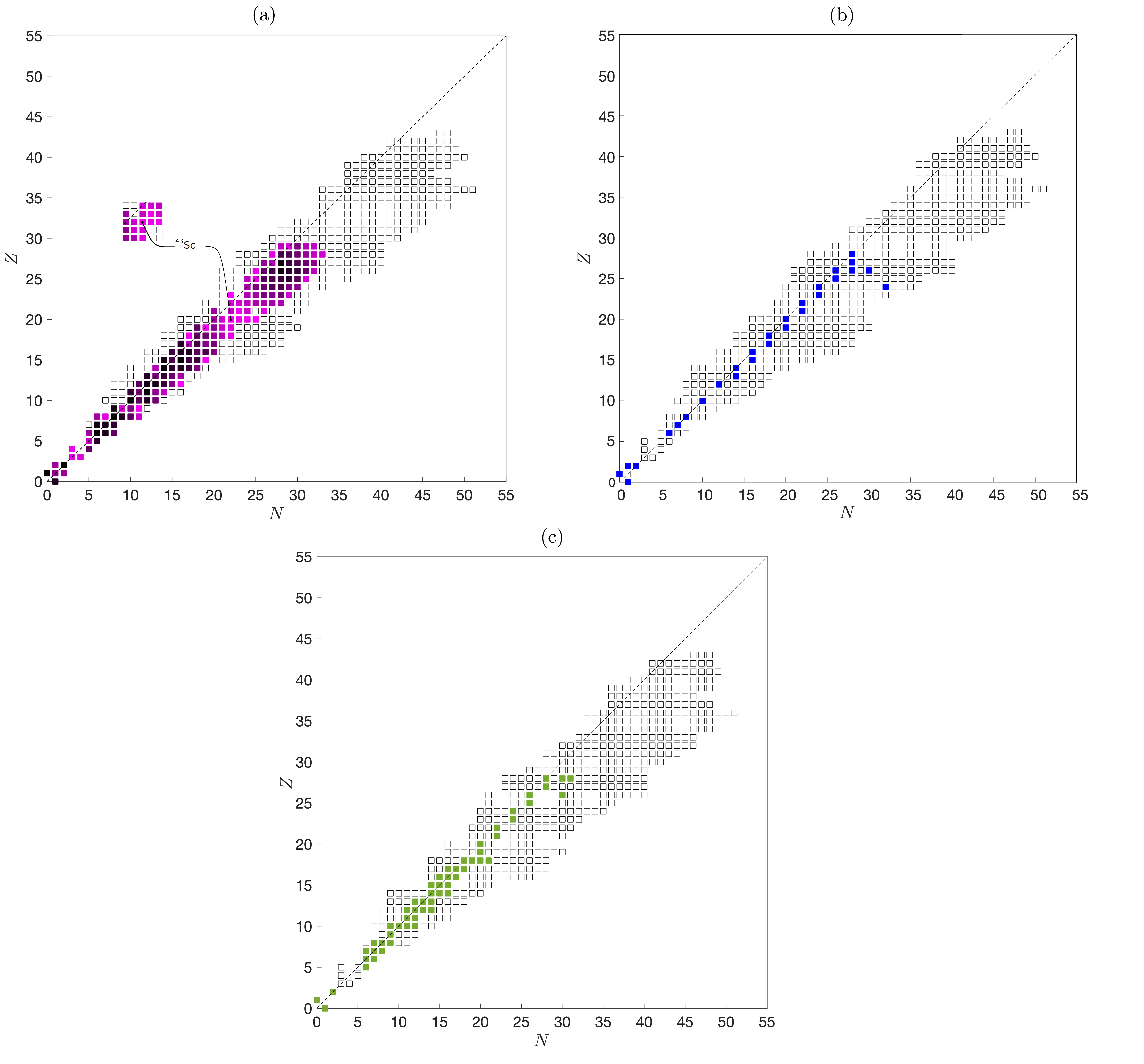,width=18.0cm}
 \caption{ (a) The first 150 ranked isotopes of Torch RN  sufficient for exact calculation of the maximum $\ce{^{56}Ni}$ abundance and energy in SNe Ia. The darker the squares, the more important (higher ranked) isotopes and the empty squares correspond to eliminated isotopes from the Torch RN. The  $^{43}\text{Sc}$ is a key isotope in  proton and neutron channels toward $^{56}\text{Ni}$ production. The f-OTD models with less than 114 isotopes (Appendix\ \ref{isotope_150}) do not contain $^{43}\text{Sc}$ and do not produce the correct amount of $^{56}\text{Ni}$.  (b) The isotopes for Approx21 and (c) SK55.}
\label{fig:Torch}
\end{figure}

A ranking is provided of the first 150 important isotopes in Appendix\ \ref{isotope_150} considering the maximum characteristic value associated with each isotope, \textit{i.e.} $w_{max,i}$. Different skeletal models can be generated by applying a threshold $\epsilon$ on  $\bw_{max}$ and eliminating isotopes and their associated reactions with $w_{max,i}<\epsilon$ from Torch RN. A comprehensive skeletal model capable of reproducing the energy and isotope mass fraction predictions of Torch RN with a certain accuracy can be developed by specifying an acceptable error level, \textit{e.g.} 5\%, in energy or $^{56}\text{Ni}$ mass fraction estimations.  Figure\ \ref{fig:Torch}(a) shows the isotopes in Torch selected for the f-OTD-150. Each square belongs to an isotope, and isotopes  with darker colors are ranked higher (selected earlier) than isotopes with lighter colors. The results show at least 114 isotopes are required to accurately produce $^{56}\text{Ni}$ with a f-OTD skeletal model. This is because of the importance of some isotopes with $21 \leq N,Z \leq 24$, and especially $^{43}\text{Sc}$ in bridging between lighter and heavier isotopes and two QSE clusters \citep{Iliadis15,SMM20}. It is shown in  Fig.\ \ref{fig:Torch}(a) that $^{43}\text{Sc}$ plays a key role for proton and neutron channels. Figures\ \ref{fig:Torch}(b) and \ref{fig:Torch}(c) show the isotopes used in the Approx21 model and SK55, a skeletal model used by \cite{TMSK19}.

\begin{figure}[!h]
\centering
 \epsfig{file=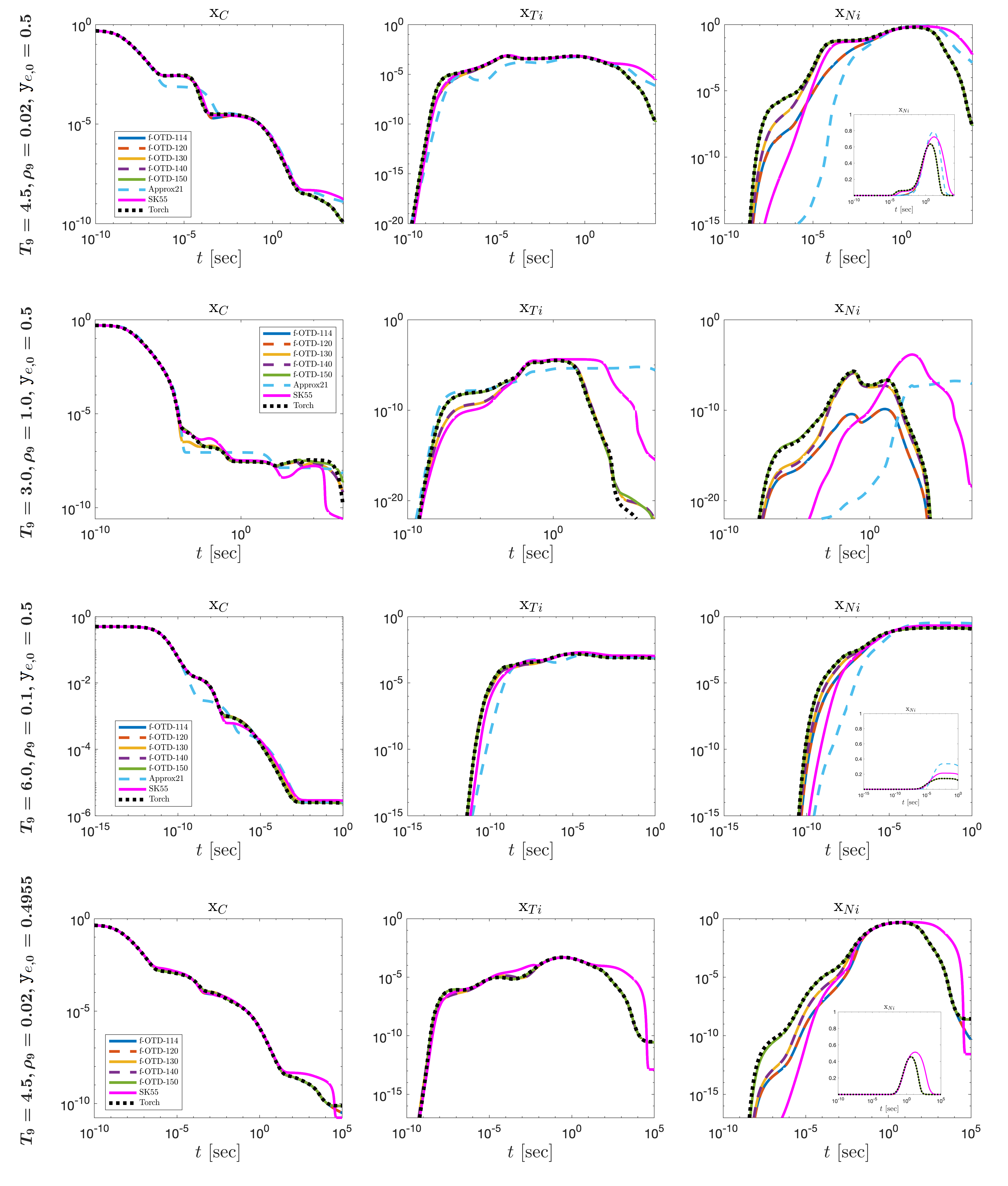,width=18.0cm}
 \caption{Model reduction on Torch RN: mass fraction estimations via Torch RN, Approx21, SK55, and f-OTD generated models for four different initial conditions of $\mathtt{x_{c12}}$, $\mathtt{x_{o16}}$, $T_9$, and $\rho_9$.}
\label{fig:Torch_massfractions}
\end{figure}

\begin{figure}[!h]
\centering
 \epsfig{file=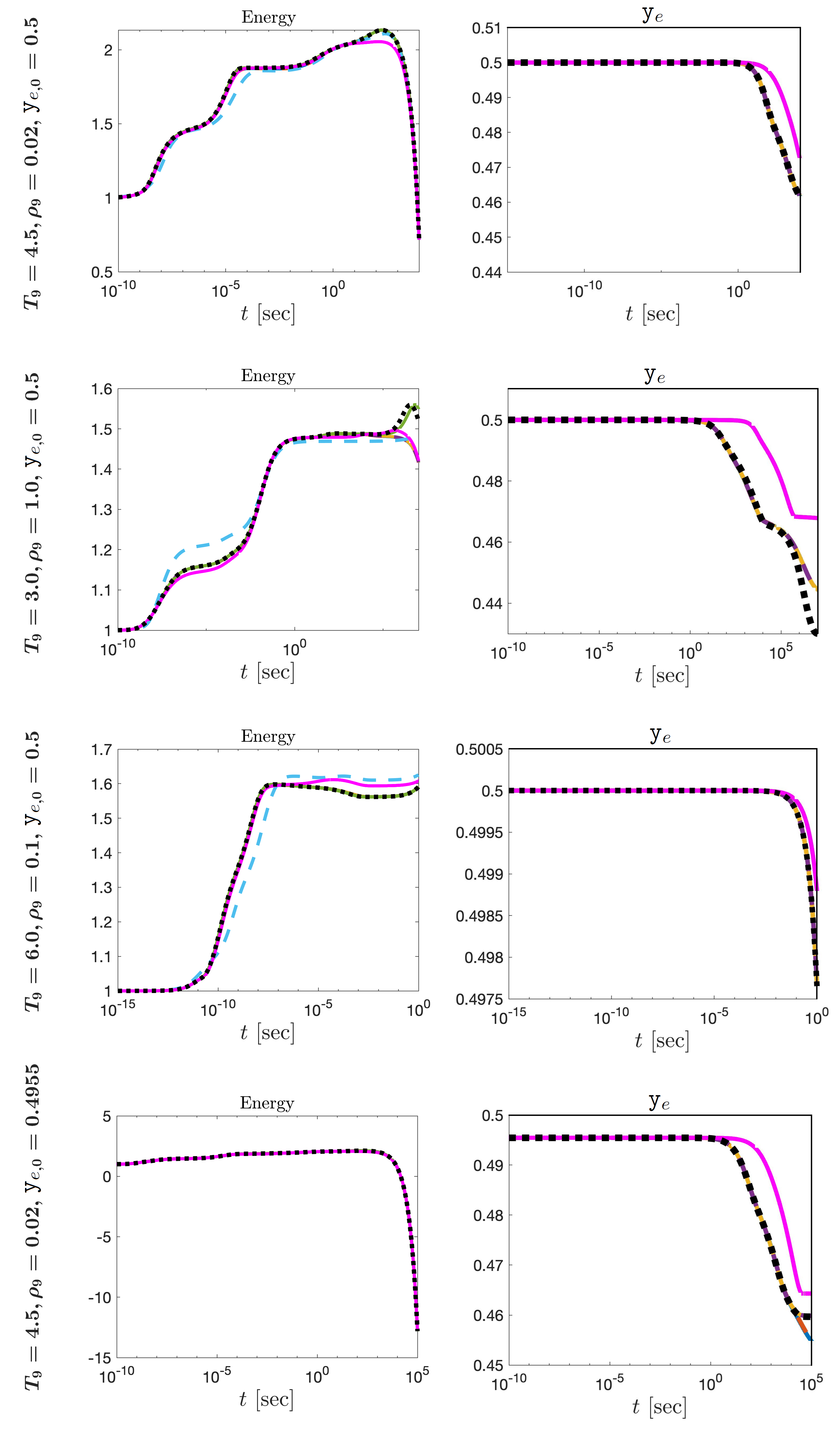,width=12.0cm}
 \caption{Model reduction on Torch RN: energy and $\mathtt{y}_e$ estimations via Torch RN, Approx21, SK55, and f-OTD generated models for four different initial conditions of $\mathtt{x_{c12}}$, $\mathtt{x_{o16}}$, $T_9$, and $\rho_9$.}
\label{fig:Torch_energyYe}
\end{figure}

\begin{figure}[!h]
\centering
 \epsfig{file=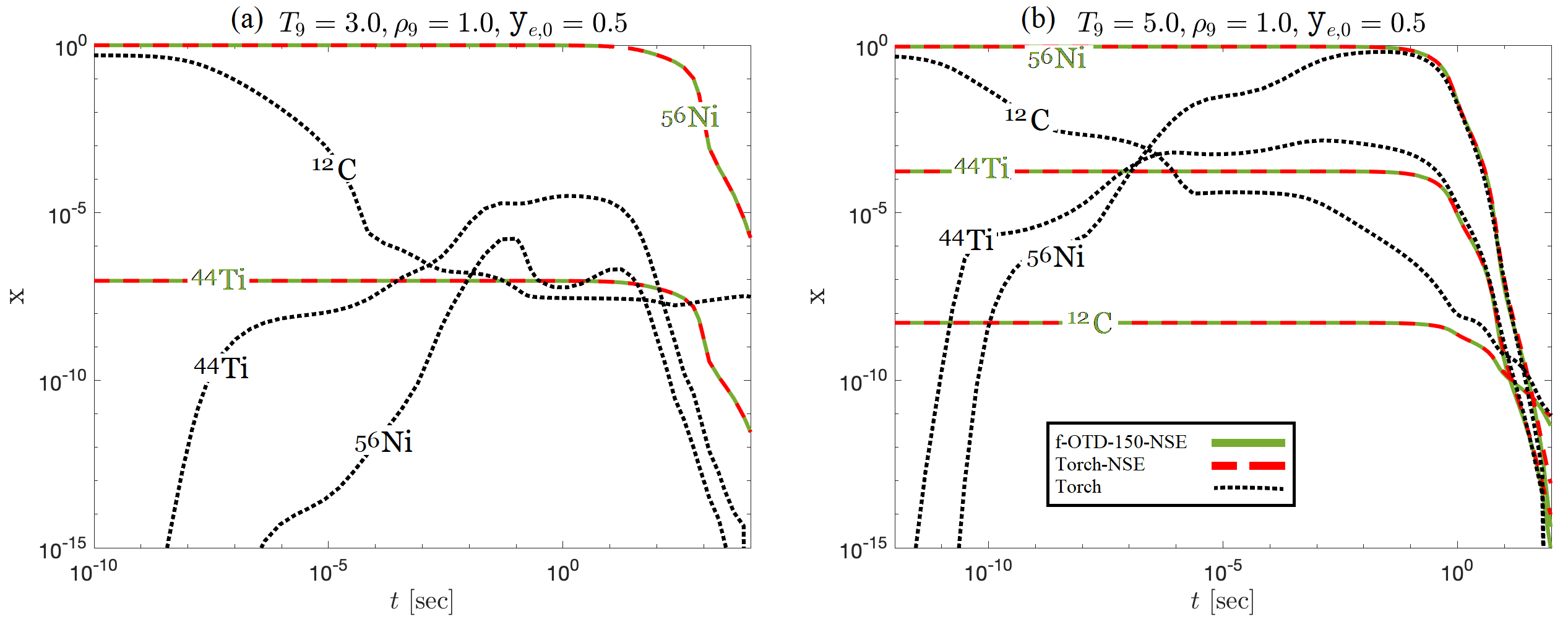,width=16.0cm}
 \caption{Comparison between mass fractions as predicted by Torch model (black dotted lines) without any equilibrium assumptions and NSE mass fractions estimated based on Torch and f-OTD-150 models (green solid and red dashed lines). NSE and non-NSE simulations have same $\mathtt{y}_e$ at each time. }
\label{fig:NSE}
\end{figure}

The Torch RN is user friendly and flexible to work with any subset of its own isotopes. This means that by providing a ranked list of isotopes from Appendix\ \ref{isotope_150}, \textit{e.g.} $\ce{^{16}O}$ (1$^{st}$ rank) to $\ce{^{45}Ca}$ (150$^{th}$ rank), one can use a f-OTD generated network with the associated isotopes. Figures\ \ref{fig:Torch_massfractions} \& \ref{fig:Torch_energyYe} show the performance of the f-OTD models in reproducing the energy, mass fractions of $^{12}\text{C}$, $^{44}\text{Ti}$, and $^{56}\text{Ni}$, and $\mathtt{y}_e$. The radioactive decays of $^{44}\text{Ti}$, and $^{56}\text{Ni}$ have significant observational applications, and the production of these two isotopes is sensitive to the temperature, density, and $\mathtt{y}_{e}$ evolution~(\cite{MTHFYW10}). The energy in Fig.\ \ref{fig:Torch_energyYe} is normalized by its initial value. The predictions via the Approx21 and SK55 are also presented. The second and third row of subfigures correspond to situations exactly similar to the test cases, but the first and last rows portray the estimations for arbitrary situations within the initial condition domain, \textit{i.e.}, $T_9 \in [2,6]$, $\rho_9 \in [0.001,1.0]$, and $\mathtt{y}_{e,0} \in [0.4955,0.5]$. It is apparent that the f-OTD models with $n_s \ge 150$  exactly predict the energy evolution of Torch RN and $^{56}\text{Ni}$ mass fraction within their designed $\rho_9$-$T_9$-$\mathtt{y}_e$ ranges. The Approx21 and SK55 models usually over-predict the maximum $^{56}\text{Ni}$ mass fraction, and Approx21 cannot be used when $\mathtt{y}_{e,0} \neq 0.5$. Replacing Torch with f-OTD models using 114-150 isotopes yields compression ratios ranging from 3.3 to 4.3. Figure\ \ref{fig:NSE} compares the evolution of mass fractions as predicted by Torch model, without any approximation, and nuclear statistical equilibrium (NSE) assumption. The NSE results are generated by using the instantaneous values of  $\rho_9$-$T_9$-$\mathtt{y}_e$ extracted from non-NSE calculations. The Torch and f-OTD-150 are used as the base reaction network for NSE estimations. It is shown in Figs.\ \ref{fig:NSE}(a) \& (b) that only at late time for $T_9=5$ the NSE and non-NSE mass fraction predictions of $^{44}\text{Ti}$ and $^{56}\text{Ni}$ (but not $^{12}\text{C}$) are close to each other.  

%As indicated earlier, a comprehensive skeletal model with good predictions covering large ranges of pressure, temperature, and mass fractions in SNe Ia should contain more isotopes than a skeletal model suitable only for a narrow range of condition. Figure\ \ref{fig:Torch_1_sample} shows that the same threshold ($\epsilon$) for producing the f-OTD-41 model out of the overall ranking in Fig.\ \ref{fig:Torch_isotope_ranking} produces a skeletal model with only 30 isotopes if only one case with $T_9=3$, $\rho_9=1$, and $\mathtt{y}_{e,0}=0.5$ is considered. The f-OTD-41 model and the 30 isotope RN show similar levels of accuracy in energy predictions for this case. 

\section{Conclusions}\label{sec:conclusion}

A systematic method for skeletal model reduction of nuclear reaction networks is developed for generating models for the carbon-oxygen combustion in SN Ia covering a range of temperatures and electron number fraction, $\mathtt{y}_{e} = Z_m /A_m$. In this method, the sensitivities of abundances with respect to reaction rates are modeled using the forced optimally time-dependent (f-OTD) method and are analyzed instantaneously. This results in reaction and isotope rankings based on the correlations between their sensitivities. A key feature of this approach is that it factorizes the sensitivity matrix into a multiplication of two low-ranked time-dependent matrices which  evolve based on evolution equations derived from the governing equations of the system. The generated skeletal models are comparatively assessed based on their ability to predict the energy and mass fractions. In particular, the skeletal models as derived here are the first to address situations covering both $\mathtt{y}_{e}=0.5$ and $\mathtt{y}_{e}<0.5$. To employ any of the skeletal models developed in this work or to create new f-OTD skeletal models with different numbers of isotopes, one only needs to feed a list of more than 114 ranked isotopes from Appendix\ \ref{isotope_150} into the Torch RN\footnote{\url{https://cococubed.com/code_pages/net_torch.shtml}}. Further reduction in the number of isotopes (\textit{e.g.}, by using equilibrium assumptions) is a potential future follow-up to this work.

The overall costs of generating an f-OTD model depend on solving the mass fraction equations (which is roughly the same as one non-f-OTD Torch simulation, $c_{torch}$), the $U$ equation, whose cost scales as $r\times c_{torch}$, and $V$ equation, whose cost scales as $r\times n_r/n_s \times c_{torch}$. For example, the total cost to generate the f-OTD-150 model with $r=1$ modes, for which 10 cases with different densities and temperatures were used, was $\sim 120 c_{torch}$, which would be negligible, for example, compared to a Monte Carlo simulation with $\sim 10,000$ trials. On the other hand, f-OTD-150 should run $\sim 3.3$ faster than Torch.

The skeletal reduction technique as described here can be readily extended to other situations. For example, lower values of $\mathtt{y}_{e}$ and higher densities to examine the production of $^{48}$Ca, $^{50}$Ti, or $^{54}$Cr. It can also be applied to more complex RNs to examine the production of heavier elements in core-collapse supernovae. With respect to the development of the methodology itself, it can be extended by including the sensitivity analysis based also on transport properties, or even the equation of state \citep{NGL19}. Most importantly, as shown recently \citep{DCB22}, the f-OTD methodology can be used for solving PDEs for multi-dimensional combustion problems in a cost-effective manner --- by exploiting the correlations between the spatiotemporal sensitivities of different species with respect to different parameters. This analysis can be especially insightful for problems containing rare events by providing more insights into global phenomena.

\section*{Acknowledgments}
This article is co-authored by D.L., an employee of Triad National Security, LLC which operates Los Alamos National Laboratory under Contract No. 89233218CNA000001 with the U.S. Department of Energy/National Nuclear Security Administration. The work at Pitt is supported by Los Alamos National Laboratory, under Contract 614709.  Additional support for the work at Pitt is provided by NSF under Grant
CBET-2042918 and Grant CBET-2152803. The authors are indebted to Prof. Frank Timmes for very useful suggestions on an earlier version of the manuscript.


%\clearpage
\vspace{2cm}
\appendix
\vspace{-0.7cm}
\section{Isotope ranking based on all cases ($\mathtt{y}_{\MakeLowercase{e},0} \leq 0.5$)} \label{isotope_150}

\begin{table}[h!]
\hspace{-2.0cm} \renewcommand{\arraystretch}{0.1}%
\begin{tabular}{ccc|ccc|ccc|ccc}
 \hline
 {Rank } & {Isotope}& {$w_{max}$ } &  {Rank } & {Isotope}& {$w_{max}$ } &  {Rank } & {Isotope}& {$w_{max}$ } &  {Rank } & {Isotope}& {$w_{max}$ }  \\ [0.5ex] 
 \hline\hline
1  & $^{16}$O  & 0.967 & 41 & $^{32}$P  & 0.112 & 81  & $^{27}$Mg & 0.020 & 121 & $^{61}$Ni & 0.003 \\
2  & $^{20}$Ne & 0.967 & 42 & $^{35}$Cl & 0.105 & 82  & $^{50}$V  & 0.018 & 122 & $^{45}$Sc & 0.002 \\
3  & $^{4}$He  & 0.967 & 43 & $^{52}$Cr & 0.094 & 83  & $^{54}$Cr & 0.018 & 123 & $^{45}$Ti & 0.002 \\
4  & $^{12}$C  & 0.935 & 44 & $^{52}$Mn & 0.094 & 84  & $^{39}$K  & 0.016 & 124 & $^{40}$K  & 0.002 \\
5  & $^{23}$Na & 0.796 & 45 & $^{56}$Ni & 0.094 & 85  & $^{24}$Na & 0.014 & 125 & $^{52}$V  & 0.002 \\
6  & p & 0.796 & 46 & $^{35}$S  & 0.093 & 86  & $^{2}$H   & 0.014 & 126 & $^{18}$F  & 0.002 \\
7  & $^{31}$P  & 0.761 & 47 & $^{57}$Co & 0.085 & 87  & $^{59}$Co & 0.012 & 127 & $^{44}$Ti & 0.002 \\
8  & $^{28}$Si & 0.743 & 48 & $^{57}$Ni & 0.085 & 88  & $^{59}$Ni & 0.012 & 128 & $^{56}$Mn & 0.001 \\
9  & $^{13}$N  & 0.732 & 49 & $^{30}$P  & 0.083 & 89  & $^{53}$Co & 0.011 & 129 & $^{58}$Cu & 0.001 \\
10 & $^{54}$Fe & 0.707 & 50 & $^{51}$Cr & 0.081 & 90  & $^{38}$Ar & 0.011 & 130 & $^{7}$Li  & 0.001 \\
11 & $^{55}$Co & 0.707 & 51 & $^{51}$Mn & 0.081 & 91  & $^{3}$H   & 0.010 & 131 & $^{41}$K  & 0.001 \\
12 & $^{17}$F  & 0.704 & 52 & $^{23}$Mg & 0.076 & 92  & $^{3}$He  & 0.010 & 132 & $^{55}$Ni & 0.001 \\
13 & $^{29}$P  & 0.681 & 53 & $^{21}$Ne & 0.075 & 93  & $^{11}$C  & 0.010 & 133 & $^{6}$Li  & 0.001 \\
14 & $^{24}$Mg & 0.669 & 54 & $^{18}$O  & 0.066 & 94  & $^{40}$Ca & 0.009 & 134 & $^{15}$O  & 0.001 \\
15 & $^{25}$Mg & 0.669 & 55 & $^{34}$S  & 0.066 & 95  & $^{41}$Sc & 0.009 & 135 & $^{40}$Ar & 0.001 \\
16 & n & 0.669 & 56 & $^{55}$Mn & 0.066 & 96  & $^{14}$O  & 0.009 & 136 & $^{47}$Sc & 0.001 \\
17 & $^{13}$C  & 0.667 & 57 & $^{9}$Be  & 0.066 & 97  & $^{48}$V  & 0.009 & 137 & $^{50}$Mn & 0.001 \\
18 & $^{14}$N  & 0.624 & 58 & $^{14}$C  & 0.061 & 98  & $^{48}$Cr & 0.009 & 138 & $^{51}$Fe & 0.001 \\
19 & $^{17}$O  & 0.624 & 59 & $^{36}$Ar & 0.052 & 99  & $^{49}$Ti & 0.008 & 139 & $^{7}$Be  & 0.001 \\
20 & $^{30}$Si & 0.518 & 60 & $^{37}$K  & 0.052 & 100 & $^{29}$Al & 0.007 & 140 & $^{35}$Ar & 0.001 \\
21 & $^{33}$S  & 0.518 & 61 & $^{57}$Fe & 0.052 & 101 & $^{50}$Ti & 0.007 & 141 & $^{43}$Ca & 0.001 \\
22 & $^{22}$Ne & 0.447 & 62 & $^{52}$Fe & 0.051 & 102 & $^{47}$Ti & 0.007 & 142 & $^{44}$Ca & 0.001 \\
23 & $^{32}$S  & 0.441 & 63 & $^{28}$Al & 0.048 & 103 & $^{47}$V  & 0.007 & 143 & $^{44}$Sc & 0.001 \\
24 & $^{33}$Cl & 0.441 & 64 & $^{37}$Ar & 0.046 & 104 & $^{48}$Ti & 0.007 & 144 & $^{34}$P  & 0.001 \\
25 & $^{25}$Al & 0.385 & 65 & $^{50}$Cr & 0.044 & 105 & $^{27}$Si & 0.006 & 145 & $^{38}$K  & 0.001 \\
26 & $^{55}$Fe & 0.352 & 66 & $^{58}$Co & 0.042 & 106 & $^{39}$Ar & 0.006 & 146 & $^{42}$Ca & 0.001 \\
27 & $^{29}$Si & 0.330 & 67 & $^{58}$Ni & 0.042 & 107 & $^{54}$Co & 0.005 & 147 & $^{20}$F  & 0.001 \\
28 & $^{11}$B  & 0.295 & 68 & $^{23}$Ne & 0.042 & 108 & $^{60}$Cu & 0.005 & 148 & $^{45}$V  & 0.001 \\
29 & $^{15}$N  & 0.285 & 69 & $^{51}$V  & 0.040 & 109 & $^{34}$Cl & 0.005 & 149 & $^{28}$Mg & 0.001 \\
30 & $^{21}$Na & 0.264 & 70 & $^{37}$Cl & 0.040 & 110 & $^{49}$Mn & 0.005 & 150 & $^{45}$Ca & 0.001 \\
31 & $^{54}$Mn & 0.191 & 71 & $^{26}$Al & 0.038 & 111 & $^{57}$Cu & 0.004 &     &      &       \\
 32 &  $^{27}$Al &  0.177 & 72 & $^{22}$Na & 0.033 & 112 & $^{58}$Fe & 0.004 &     &      &       \\
33 & $^{56}$Fe & 0.163 & 73 & $^{49}$V  & 0.030 & 113 & $^{46}$Ti & 0.004 &     &      &       \\
34 & $^{56}$Co & 0.163 & 74 & $^{49}$Cr & 0.030 & \cellcolor[gray]{0.9} 114 & \cellcolor[gray]{0.9}  $^{43}$Sc & \cellcolor[gray]{0.9} 0.004 &     &      &       \\
35 & $^{26}$Mg & 0.161 & 75 & $^{59}$Cu & 0.027 & 115 & $^{10}$B  & 0.004 &     &      &       \\
36 & $^{33}$P  & 0.155 & 76 & $^{53}$Cr & 0.027 & 116 & $^{60}$Ni & 0.003 &     &      &       \\
37 & $^{31}$Si & 0.142 & 77 & $^{19}$F  & 0.025 & 117 & $^{61}$Cu & 0.003 &     &      &       \\
38 & $^{53}$Mn & 0.128 & 78 & $^{32}$Si & 0.024 & 118 & $^{25}$Na & 0.003 &     &      &       \\
39 & $^{53}$Fe & 0.128 & 79 & $^{36}$S  & 0.022 & 119 & $^{19}$O  & 0.003 &     &      &       \\
40 & $^{31}$S  & 0.124 & 80 & $^{36}$Cl & 0.022 & 120 & $^{41}$Ca & 0.003 &     &      &      
\\ [0.5ex] 
 \hline\hline
\end{tabular}
\end{table}

 \clearpage
\FloatBarrier
\bibliography{SN,ROM_combustion,HB,cfd_arash}
\bibliographystyle{aasjournal}



\end{document}